\documentstyle{cupconf}

\newcommand{\be}{\begin{equation}}
\newcommand{\ee}{\end{equation}}
\newcommand{\ba}{\begin{eqnarray}}
\newcommand{\ea}{\end{eqnarray}}

\ifoldfss
\else
  \ifnfssone
    \newmathalphabet{\mathit}
      \addtoversion{normal}{\mathit}{cmr}{m}{it}
      \addtoversion{bold}{\mathit}{cmr}{bx}{it}
    \newmathalphabet{\mathcal}
      \addtoversion{normal}{\mathcal}{cmsy}{m}{n}
    \else
    \ifnfsstwo
    \fi
  \fi
\fi

\def\hexnumber#1{\ifcase#1 0\or1\or2\or3\or4\or5\or6\or7\or8\or9\or
 A\or B\or C\or D\or E\or F\fi }

\makeatletter
\ifx\CUP@mtlplain@loaded\undefined
\else
\fi
\makeatother
 \makeatletter
 \ifx\CUP@mtlplain@loaded\undefined
   \font\tenbmi=cmmib10 at 10pt
   \font\sevenbmi=cmmib10 at 7pt
   \font\fivebmi=cmmib10 at 5pt

   \newfam\bmifam
   \textfont\bmifam=\tenbmi
   \scriptfont\bmifam=\sevenbmi
   \scriptscriptfont\bmifam=\fivebmi
   
 \fi
 \makeatother
\ifnfsstwo

\fi
\ifnfssone

\fi
\ifoldfss

\fi

\mathchardef\varLambda="0103

\makeatletter
\ifx\CUP@mtlplain@loaded\undefined
\else
\fi
\makeatother
\makeatletter
\ifx\CUP@mtlplain@loaded\undefined
  \font\tenbms=cmbsy10
  \font\sevenbms=cmbsy10 at 7pt
  \font\fivebms=cmbsy10 at 5pt
  \newfam\bmsfam
  \textfont\bmsfam=\tenbms
  \scriptfont\bmsfam=\sevenbms
  \scriptscriptfont\bmsfam=\fivebms

  \edef\bsy@{\hexnumber\bmsfam}
  \mathchardef\bnabla="0\bsy@72
\fi
\makeatother
\title[Turbulence in HI]{Turbulence in Atomic Hydrogen}

\author[A. Lazarian]%
{A.\ns L\ls A\ls Z\ls A\ls R\ls I\ls A\ls N\thanks{e-mail:
lazarian@selene.princeton.edu}}

\affiliation{ Princeton University Observatory, Princeton NJ 08544}

\begin{document}
\ifnfssone
\else
  \ifnfsstwo
  \else
    \ifoldfss
      \let\mathcal\cal
      \let\mathrm\rm
      \let\mathsf\sf
    \fi
  \fi
\fi

\maketitle

\begin{abstract}
Understanding the properties of interstellar turbulence is 
a great intellectual challenge and the urge to solve this problem
is partially motivated by a necessity to explain the star formation
mystery. This review deals with a recently suggested inversion technique 
as applied to atomic hydrogen. This technique allows to determine 3D
turbulence statistics through the variations of 21~cm intensity. We claim
that a radio interferometer is an ideal tool for such a study as its 
visibility function is directly related to the statistics of galactic HI. 
Next, we show how galactic rotation curve can be used to study the turbulence
slice by slice and  relate the statistics given in galactic coordinates 
and in the velocity space. The application of the technique to HI data 
reveals a shallow spectrum of the underlying HI density that is not 
compatible with a naive Kolmogorov picture. We show that the random density 
corresponding to the found spectrum tends to form low contrast filaments 
that are elongated towards the observer. 
\end{abstract}

\firstsection 
\section{Introduction}

The properties of the interstellar medium strongly suggest that it is
turbulent. Here turbulence is understood  as unpredictable 
temporal behavior of nonlinear systems as preached by J. Scalo (1985,
1987). 

The importance of turbulence in molecular clouds and its relation
to star formation has long been appreciated (Dickman 1985). Recent
progress in numerical simulations of molecular cloud dynamics
(see Ostriker, this volume) indicates the intrinsic connection
between the turbulence in different phases of the interstellar medium  
(McKee \&  Ostriker 1977).
In what follows we shall mostly discuss the turbulence in atomic hydrogen (HI),
although the formalism presented here is applicable to other spectral lines.

Statistical description is a nearly indispensable strategy when
dealing with turbulence and a big advantage of statistical techniques is
that they extract underlying regularities of the flow and reject
incidental details. 
Kolmogorov notion of energy cascade from large to small scales
has been proved an extremely
valuable concept and Kolmogorov spectrum of turbulence
has been measured in various media.    At
the same time, astrophysical turbulence, unlike that in incompressible
fluids, is a much more complicated
phenomenon, and therefore one cannot {\it a priori} hope that 
Kolmogorov's (1941) description is adequate
(cf. Armstrong  et al.  1995). Energy injection at small scales,
shocks, compressibility may make interstellar turbulence spectrum
much more informative, and we should expect to see deviations
from the boring $-11/3$ slope.

The advantage of using 21~cm emission data  is 
that a continuum of separations between data points is available.
This property is shared by  diffuse emission in other spectral
lines, but 21~cm measurements allow to disregard dust adsorption.
As our review deals with HI studies within the galactic
plane we do not discuss in detail interesting results obtained
for HI in Large Magelanic Clouds (Spicker \& Feitzinger 1988a,b).
To avoid possible misunderstanding we should stress that Spicker
 \& Feitzinger (1988a,b) deal with velocity fluctuations,
while only intensity fluctuations are available when one
studies HI turbulence in galactic plane. Statistics
of random velocity and density fields may be different
and therefore a direct comparison of the results obtained
for these fields may be misleading. 

Being limited in space
we refer the interested reader to the earlier reviews on
interstellar turbulence, among which the one by Dickman (1985) 
can serve as an excellent introduction to the basic statistical
techniques. A more advanced reader will enjoy a thoughtful
analysis of problems associated with the statistical analysis
of observational data given in Houlahan \& Scalo (1990).  
Important aspects of the statistical analysis are discussed,
for instance, by Dickman \& Kleiner (1985), Roy \& Joncas (1985)
Perault  et al.  (1986), 
O'Dell \& Casta\~{n}eda (1987),  Rickett (1988),
 Van Langevelde 
et al. (1992)  Kitamura  et al.  (1993), 
Meisch \& Bally (1994), Armstrong, Rickett, \& Spangler (1995),
Wallin, Watson, \& Wyld (1998) and by the contributors
to the present volume. A brief discussion of the very early
studies of interstellar statistics can be found in Lazarian (1992).

Studies of interstellar turbulence frequently deal with samples 
which are not statistically homogeneous (see Miesch \& Bally 1994).
Indeed, whenever individual molecular complexes are studied, 
the statistics (especially at large separations)
may be dominated by regular gradients rather than the random 
component. To eliminate the inhomogeneous component,
various types of spatial filtering are used (see  Zurfleh 1967). 
These problems are
alleviated for HI studies, since molecular clouds tend to
be localized objects in sharp contrast to more pervasive distribution
of atomic hydrogen.

Further on we shall deal with the two point structure functions
and power spectra. Naturally, one cannot place
pickup devices at different points of HI. Instead, only the 21-cm
intensity fluctuations with pointwise emissivity integrated along the
lines of sight are available. It is obvious, that given a statistical
description of the transparent emitting astrophysical medium, it is
possible to predict statistical properties of the observable diffuse
emission (Kaplan \& Pikelner 1970), which would correspond
to the solution of the {\it forward problem}. However, more important
is to solve the {\it inverse problem}, namely,
to deduce the 3D statistics of HI from observations. These issues
are dealt with in sections 2 and 3. In section 4 we discuss 
the application of the technique to interferometric data.

The galactic rotation curve allows
one to study turbulence slice by slice. This slicing, however, is far from
trivial (see section 4). Indeed, topologically disconnected blobs
of HI can overlap in the velocity space if their velocities are
the same. We show that 
interferometric study can potentially provide the information
about both random density and velocity fields.

Addressing the issue of HI topology we show that HI with the measured
spectrum of density fluctuations forms low contrast filaments
(section 5) and these filaments are elongated towards the observer due
to the presence of velocity fluctuations.

\section{Intensity and Density Fluctuations}

The 21~cm intensity
of in ${\bf e}_{i}$ direction can be presented as
\begin{equation}
I({\bf e}_{i},\omega, \bigtriangleup\omega)=
\int^{\omega_{2}}_{\omega_{1}}
i({\bf e}_{i}, \omega)\,d\omega~~~,
\end{equation}
where $i({\bf e}_{i}, \omega)$ is the emissivity at the
central frequency $\omega$ inside the bandwidth $\bigtriangleup
\omega=\omega_{2}-\omega_{1}$. Expressing the emission intensity
through density, it is easy to obtain in the case of an `optically
thin' gas (Spitzer 1968)
\begin{equation}
I({\bf e}_{i},\omega, \bigtriangleup\omega)\sim \int_{0}^{L}
n(x,{\bf e}_{i}){\rm d}x~~~,
\label{7}
\end{equation}
where $n(x,{\rm \bf e}_{i})$ is
the number of atoms emitting along the $x$-axis in the bandwidth
$\bigtriangleup \omega$, and the integration is performed along the
line of sight.

As the number of atoms along different lines of sight varies, the
intensity of HI emission fluctuates. These fluctuations can
be characterized via structure functions of intensity\footnote{In Eq.~(\ref{eqn.3}) and further
on it is assumed that the $\bigtriangleup \omega$ and the frequency
$\omega$ are fixed.} 
\begin{equation}
D({\rm\bf e}_{1}, {\bf e}_{2}) =
\langle (I({\bf e}_{1})-
I({\bf e}_{2}))^{2}\rangle~~~.
\label{eqn.3}
\end{equation}

Since the sky is observed as
a two-dimensional manifold, two coordinates characterize the relative
position of two correlating points. As these coordinates we use
the angular separation between the points $\theta$ ($\cos\theta=
\frac{{\bf e}_{1}\cdot{\bf e}_{2}}{|{\rm \bf e}_{1}|\cdot
|{\rm \bf e}|}$) and the positional angle $\varphi$.  
  Then, the structure function of intensity 
$D({\bf e}_{1}, {\bf e}_{2})=D(\theta,\varphi, \bf E)$,
where a two dimensional vector $\bf E$ characterizes the turbulent 
volume under study. 
It is easy to see that 
the function $D(\theta,\varphi, \bf E)$ can be used to crudely estimate
the correlation scale of the turbulence and its $\varphi$ dependence
characterizes the turbulence anisotropy. 

For a more quantitative treatment one has to relate the statistics
of HI fluctuations to the statistics of the underlying turbulence.
Generalizing the treatment suggested in Kaplan \& 
Pickelner (1970) we formulate an integral
equation for the {\it forward problem} (see Lazarian 1995, henceforth L95):
\begin{eqnarray}
D_{0}(p,\varphi,{\bf E})
&=&
\mbox{\ae}^{2}\int\!\!\int_{0}^{L}\{
{\rm d}(\sqrt{(x_{1}-x_{2})^{2}+p^2}, 
\varphi,\psi ,\tau,{\bf E})\nonumber\\
&-&
{\rm d}(|x_{1}-x_{2}|, \tau,{\bf E})\} dx_{1} dx_{2}~~~,
\label{e.5}
\end{eqnarray}
where $\tau=(x_1+x_2)/2$ and the structure function of density
\begin{equation}
{\rm d}(\sqrt{(x_{1}-x_{2})^{2}+p^2}, 
\varphi,\psi ,\tau,{\bf E}) = 
\langle(n(x_{1},{\bf e}_{1}) - n(x_{2},{\bf e}_{2}))^{2}\rangle~~~,
\end{equation} 
depends on 6 coordinates, among which $\psi$ is 
the angle
between the radius vector connecting the correlating points and the line of
sight (see Fig.~1).

\begin{figure}
\vspace{7cm}
\caption{The schematic of lines of sight and the region under study.
The slice of HI with the thickness $L$ is observed from such a large
distance $R$ that lines of sight ${\bf e}_1$ and ${\bf e}_2$ 
are nearly parallel within the slice. Various turbulence scales, e.g. $l_1$,
$l_2$, contribute to the correlation function of density for
the fixed $p=R\theta$. $l_1$ and $l_2$ make angles $\psi_1$ and $\psi_2$
with the lines of sight. Therefore the dependence of structure functions
on $\psi$ is important for the inversion. For the isotropic density field
structure functions do not depend on $\psi$.}
\end{figure}

To solve the {\it inverse problem}, namely, to find the statistics
of turbulence via the statistics of intensity fluctuations one has to
make some assumptions about the turbulence. The
assumption of local isotropy of turbulence 
was tested by Green (1994), who found no appreciable anisotropy in 
his data. Therefore  we can follow a simplified version of the
treatment suggested in L95 (see also Lazarian 1994a)
and disregard $\varphi$ and $\psi$ 
dependences
of the density structure functions. As a result we get (L95): 
\begin{equation}
\int_{0}^{L}\{{\rm d}(r,\tau)-{\rm d}(L_{1},\tau)\}{\rm d}\tau\approx
-\frac{1}{\pi}
\int_{r}^{L}
\frac{{\rm d}p}{\sqrt{p^{2}-r^{2}}}D'(p)~~~. 
\label{e.18}
\end{equation}
It is a common practice in turbulence literature
to separate the dependencies on $\tau$ and 
$r$ variables and present the structure function 
as a product of two functions (see Isimary 1978): $ {\rm d}(r,\tau)={\rm d}(r)\cdot C(\tau)$. 
Then, the integral over $\tau$ in Eq.~(\ref{e.18}) gives only a 
scaling constant $C_{1}$. If $C(\tau)$ changes slowly over the interval 
$L$, $C_{1}\sim L$.

The physical meaning of the inversion above can be  easily understood.
As we can see in Fig.~1, various scales of turbulence from $p=R\theta$
to some maximal cutoff
scale contribute to the structure functions of intensity $D(p)$. At a 
different angular separation 
$p'=R\theta'>p$ scales from $p'$ to the same cutoff contribute to the
structure function 
 $D(p')$. It is obvious, therefore, that the difference
$D(p')-D(p)$ contains the information about the turbulent
scales from $p$ to $p'$.

\section{Velocity Fluctuations}

The inversion procedure above is rigorous only when velocity fluctuations
can be disregarded. For HI studies in the galactic plane
velocity fluctuations should be accounted for. Indeed, even in 
the absence of density inhomogeneities, intensity fluctuations
in the velocity space can be  produced by random
velocity. Moreover, slicing of hydrogen may become ambiguous.
We may recall, that the slicing assumes a monotonic dependence of 
velocity on  distance. The random velocity  $u^{turb}$
distorts HI motion, which otherwise would be determined by
the Galactic rotation curve. The latter motion is characterized
by the projection of the regular velocity  to the line of sight
(z-axis) $V^{reg}$  and its spatial derivative
$f=(\delta V^{reg}/\delta z)$. Since the actual velocity along
the line of sight is $ V^{reg}+u^{turb}$ spatially distant regions may 
be mapped into the same
slice, while adjacent regions with different velocities will be mapped
into different slices. It is also obvious that the turbulence
statistics in the velocity space is anisotropic even if the statistics
is isotropic in galactic coordinates. Indeed, only the velocity
component along the line of sight matters and this makes
the direction towards the observer ``the chosen direction''. 

One may wander whether the statistical
treatment described in section 2 is applicable to atomic hydrogen
in the Galactic plane. Obviously, our analysis is not sensitive
to velocity fluctuations when the integration over the whole 
21~cm
line is performed. It is also intuitively clear that when the
thickness of the HI slice in the velocity space $\triangle V$ is much larger
than the turbulent velocity dispersion $\delta u_k$ 
at the scale under study, the
fluctuations of velocity are marginally important. The quantitative
treatment in Lazarian \& Pogosyan (1998) (henceforth LP98) proves this.
At the same time for $\triangle V< \delta u_k$
the velocity fluctuations may dominate the measurements.

To  distinguish the cases when velocity fluctuations are important
and negligible it is convenient to talk about ``thick'' 
and ``thin'' slicing of data. Using this terminology we may say
 that L95, where velocity
fluctuations were disregarded, dealt entirely with ``thick'' slicing.

We may show\footnote{A rigorous treatment is given in LP98, while here
we present simplified  estimates.} the difference between the ``thin'' 
and ``thick'' slicings 
assuming that
the Fourier modes of density are independent
random numbers  in the velocity space. The density at point 
$({\bf P}, v_z)$, where
$\bf P$ is a two dimensional vector in $xy$ plane, is
\be
\delta n({\bf P}, v_z)\sim \int d{\bf K} dk_z F^{1/2}({\bf K}, k_z) 
\exp(i{\bf K}{\bf P})
\exp(ik_z v_z f)~~~,
\ee
where $F({\bf K}, k_z)$ is the underlying 3D spectrum of HI random density.
As 21~cm intensity is proportional to the integral of $\delta n$ over
the thickness of the velocity slice, the correlation
function of intensity is
\be
\langle \delta I({\bf P}) \delta I({\bf P_1})\rangle\sim
\int_{\delta v_z} d~v_{z1} d~v_{z2}
\langle \delta n({\bf P}, v_{z1}) \delta n({\bf P_1}, v_{z2})\rangle
\sim \int d{\bf K} \exp[i{\bf K}({\bf P}-{\bf P_1})]F_2({\bf K})
\ee
where $ \delta I$ and $ \delta n$ are
variations of intensity and density, respectively, while
 the two dimensional spectrum $F_2({\bf K})$ is given by
\be
F_2({\bf K})\sim \int dk_z
\int_{0}^{\triangle V} d~v_z \exp[ik_z v_z f] (1-v_z/\triangle V) 
F({\bf K}, k_z)~~~,
\label{slice}
\ee
where $\triangle V=|v_{z1}-v_{z2}|$ is the thickness of HI slice in the 
velocity space.

First consider {\it ``thick'' slicing} $|{\bf K}|\gg 1/f\bigtriangleup V$. The contribution
to the integral (\ref{slice}) comes mostly from $k_z < 1/f\triangle V$,
as for larger $k_z$ the exponent oscillates rapidly and the inner
integral in Eq.~(\ref{slice}) is small. Therefore
\be
F_2({\bf K})\sim  F({\bf K}, \bar{k}_z)~~~,
\label{thick}
\ee
where $\bar{k}_z$ is a value in the interval 
$[0, 1/f\bigtriangleup V]$.
If the turbulence is isotropic\footnote{The turbulence is definitely
anisotropic in the velocity space. However the underlying turbulence 
may be isotropic in galactic coordinates. Eq.~(\ref{thick}) shows
that $P_2$ is mostly determined by fluctuations in $xy$ plane where
no velocity distortions are present.} its spectrum 
 in galactic coordinates, 
 $F({\bf K}, \bar{k}_z)=
F(\sqrt{|{\bf K}|^2+\bar{k}_z^2})\approx F(|{\bf K}|)$, 
and $F_2(|{\bf K}|)\sim  F(|{\bf K}|)$ in agreement with L95.

In the case of {\it ``thin'' slicing} $|{\bf K}|\ll 1/f \triangle V$, 
 for most $k_z$
$\exp[ik_z v_{z}f]\approx 1$ 
and 
\be
F_2({\bf K})\sim  \triangle V \int dk_z F({\bf K}, k_z)~~~.
\label{thin}
\ee

The spectrum (\ref{thin}) does depend on the distortions introduced by velocity
fluctuations. Therefore thin slicing can provide the information
about the velocity spectrum. This issue is studied in LP98 where the
spectrum of density in the velocity space is derived.

\section{Interferometric Studies and Spectrum of Galactic HI}

The theory of interferometric study of HI turbulence was discussed in L95. 
There it was found that the sum of the sine and cosine parts of
the interferometer visibility function is proportional to the spectrum
of the locally isotropic HI random density field. This result can be easily
understood
in view of our earlier discussion (see section~3). Indeed, using an
interferometer one can get a two dimensional spectrum of 21~cm intensity;
Eq.~(\ref{thick}) relates this spectrum 
(in the case of ``thick'' slicing) 
to the
underlying spectrum of density fluctuations.

The application of the technique to Green's 
(1993) 
DRAO data towards $l= 140^{\circ}, 
b=0^{\circ}$ ($03^{h} 03^{m} 23^{s}, +58^{\circ} 06' 20'$, epoch 1950.0)
reveals a shallow spectrum of density with the power-law index 
$\sim -3$ (L95), compare to $-11/3$ for the Kolmogorov turbulence\footnote{L95
uses the spectrum integrated over directions in $k$-space. For such a
spectrum
the Kolmogorov turbulence has the index $-5/3$ and the measured spectrum 
of HI density has the index $\sim -1$.}. According to a simple
flat rotation curve the slicing of observational
data  corresponded to $\approx 2.2$~kpc for the most  distant HI
and $\approx 0.6$~kpc for the closer HI. This should be compared with
largest turbulence scales $\approx 170$~pc and $\approx 40$~pc
that were studied for distant and closer regions, respectively.
Adjacent channels of the interferometer had central velocities
separated by 5~km/s, which is larger than the random velocity dispersion 
for most of the scales studied\footnote{Wherever this is not true,
several channels should be combined to obtain ``thick'' slicing of data.}. 
Therefore the slicing is ``thick'' which justifies the use of L95 technique.
The application of the technique to inferior quality data in Crovisier
\& Dickey (1983), Kalberla \& Mebold (1983),
Kalbela \& Stenholm (1983) also suggests a shallow spectrum of turbulence, 
but the 
accuracy of determining the spectral index is poor.

If interferometers with higher velocity resolution are used, one
can use both ``thick'' and ``thin'' slicing. ``Thick'' slicing
can provide the information about the density spectrum and its
combination with the ``thin'' slicing can provide the information
about the velocity field (see Eq.~(\ref{thin})). For HI blobs out of galactic
plane the velocity spectrum can be directly determined (Lazarian 1994b).

\section{HI Filaments}

An issue closely related to the spectrum of atomic hydrogen is the statistical
formation of HI filaments. We talk about  statistical origin as opposed
to  dynamical formation. Although any particular  
enhancements of density
do have dynamical origin, Lazarian \& Pogosyan 1997 (henceforth LP97)
have shown that large scale filaments may appear in a medium without the 
action of forces correlated on large scales. Indeed, the major result
of LP97 is that low contrast filaments are formed naturally in 
a Gaussian density field\footnote{Filaments in other random density fields, 
e.g. log-normal density field, are discussed in LP98.}  when the power 
spectrum corresponds to observations. Statistical formation
may explain, for example, mysterious filaments observed by Verschuur (1991a,b).

In the presence of velocity fluctuations, LP98 show that the filaments
become elongated and directed towards the observer. This is some sort
of the ``fingers of God'' effect\footnote{The analogy is not
exact as turbulence is characterized by a power law spectrum of
fluctuations as opposed to a boring exponential distribution of
density in clusters of galaxies. Therefore interstellar ``fingers''
are not trivial.} (see Peebles 1971)
that is well known  in the studies of clusters of galaxies.  Figure~2 shows
isocontours of density at one $\sigma$ level ($\sigma$ is the dispersion
of density fluctuations) after smoothing on 
the scale $0.03$ of the computational box size. It is assumed that
the velocity field is Kolmogorov, while the density field has a
spectral index $-2.7$, which is close to that observed in L95.

\begin{figure}
\vspace{11 cm}
\caption{Filaments at large scales corresponding to the 
distribution of random density with the power law index 
$-2.7$ in the presence of velocity fluctuations with
Kolmogorov spectrum (from LP98). The 
upper left 
panel  shows a projection in x-y plane and it looks very similar to
pictures in LP97. Indeed, the spectrum in this plane is
not distorted by space-velocity mapping. On the contrary, the distribution
of density in the upper right panel is anisotropic. 
The complex structure
of the filaments is more vivid in the lower right panel 
where a zoomed part of the upper right panel is shown. 
The 3D structure of filaments in the velocity
space is shown in the lower left panel.}
\end{figure} 

A study in LP98 has shown that for scales larger than $\sim 250$~pc the
velocity distortion of the structures become marginal and therefore
statistically formed filaments should be  distributed uniformly,
in correspondence with figures in LP97.

It worth noting, that filaments discussed are real physical entities and
not an artifact of data handling. Therefore a 
tight correlation in gas
and dust emission is expected, which corresponds to the results in
Verschuur (1994). 
In other words,
we expect that filaments observed in 21~cm will be also seen at 100$\mu$.

\section{Future Work}

\subsection{Inversion and Diffuse emission}

Extensive studies of HI statistics are on the agenda. It is challenging,
for instance,
to compare statistical properties of HI in the regions with different
rates of star formation.

The inversion technique discussed is widely applicable.
Not only 21~cm emission but diffuse emission in lines and
synchrotron can be used.
In fact, a statistical technique that
allows studying Galactic regular and random magnetic fields via fluctuations of
the synchrotron intensity predated the development of HI inversion technique
(see Lazarian \& Shutenkov 1990, Lazarian 1992). Techniques that use 
variations of
other Stock's parameters are under development (Spangler, private 
communication). We believe that when techniques that 
account for extinction are developed, the inversion
 will be used for various spectral lines. 
Numerical techniques with many-frequency input
are likely to be called for and the turbulence inversion will become 
as informative for molecular clouds as helioseismology is for the
Sun (Gough 1985).

\subsection{Other Approaches}

Power spectrum and two point structure functions are the statistical
tools most frequently used in turbulence studies. They are not the only
ones, however. Numerous studies show the advantage of using wavelet transform
techniques (see Gill \& Henriksen 1990, Langer, Wilson, \& Anderson 1993) 
and one point statistics (Semadeni, this volume). N-point statistics
have been successfully used in the studies of Large Scale Structure
(Peebles 1980), but they still must make their way to the domain of
interstellar turbulence and, in particular, HI turbulence studies.
Some of the statistical tools described in Peebles (1980)
are remarkable informative, for instance, they can distinguish between
collapsing and expanding blobs of gas.

Genus statistics (see Gott et al 1989) seems to be a promising tool for 
determining HI
topology. This statistics allows to distinguish between ``sponge-like''
topology and ``meat ball'' topology of atomic hydrogen. Possible
variations
of HI topology from region to region may be related to the variations
of the filling factor of the cold neutral medium. 

\section{Summary} 

The following are the principal conclusions of this review:

\begin{itemize}
\item The application of a newly developed inversion technique reveals
a shallow spectrum of HI density in galactic plane. The spectrum is different
from the Kolmogorov one, which may be indicative of non-trivial physics 
involved.

\item The spectrum of HI density is anisotropic in velocity space with
velocity fluctuations altering the statistics along the line of sight.
The degree of anisotropy and the spectrum in velocity space 
depend on the scale under study. The spectrum become isotropic for scales
larger than 200-300~pc.

\item Interferometers are useful tools for studying HI turbulence 
in galactic plane and
may provide the spectrum of 3D random density field
and useful information on random velocity field.

\item Atomic hydrogen with Gaussian density and the shallow spectrum
corresponding to observations forms low contrast
filaments. These filaments are anisotropic in the velocity space and
directed towards an observer.
\end{itemize}

\begin{acknowledgments}
I would like to acknowledge many useful discussions with B.~Draine,
J.~Cordes, S.~Spangler. I especially value comments by D.~Gough
on the inversion procedure. My fruitful collaboration with D.~Pogosyan
is happily acknowledged.  The research is
supported by NASA grant NAG5 2858.
\end{acknowledgments}

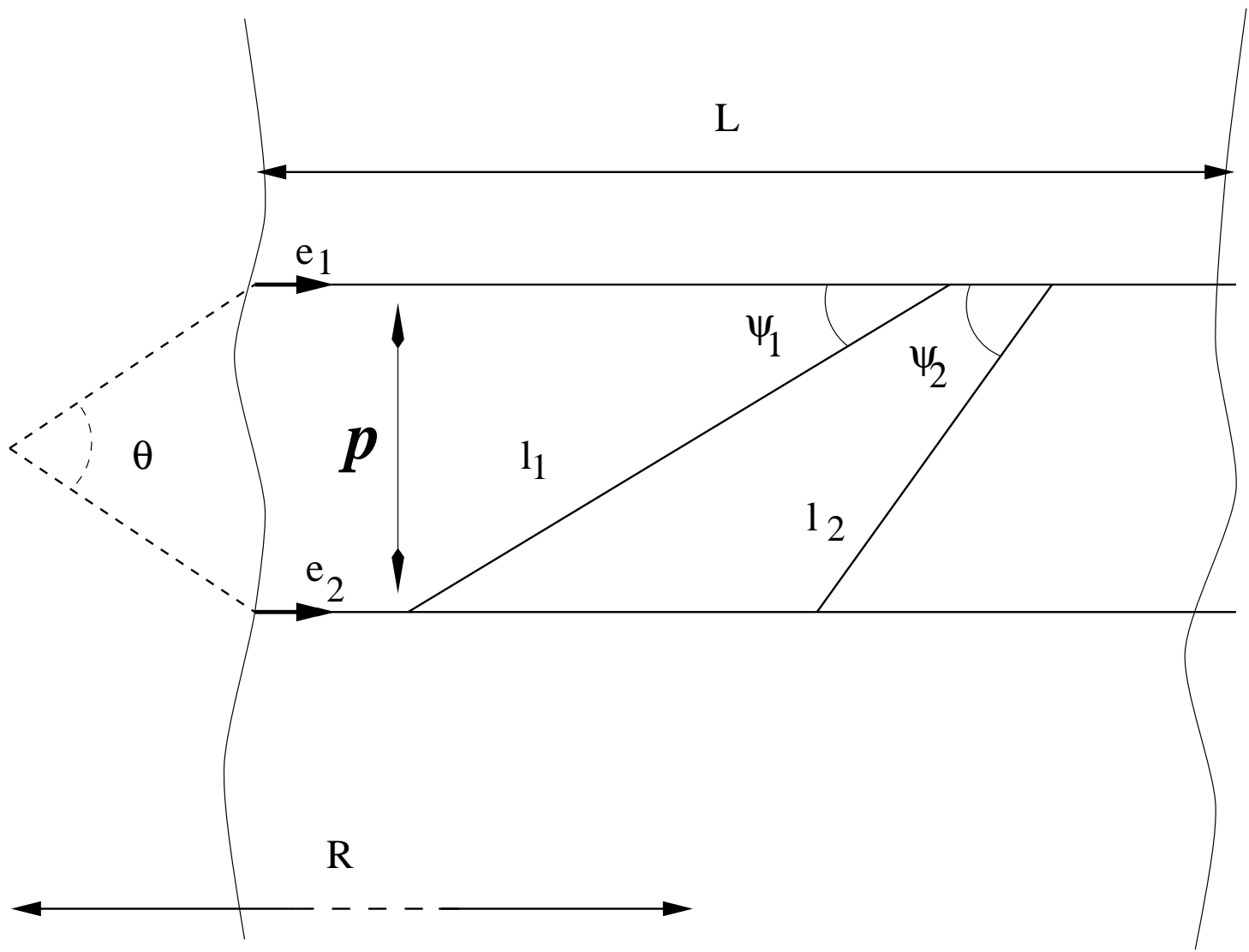
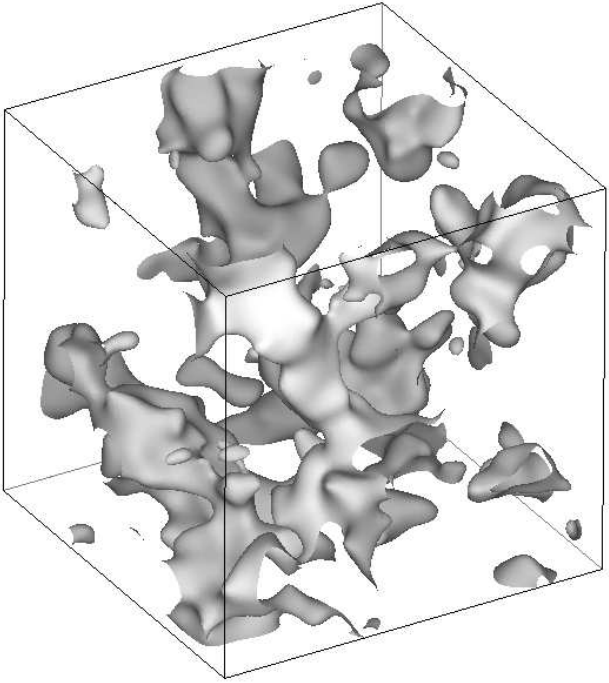
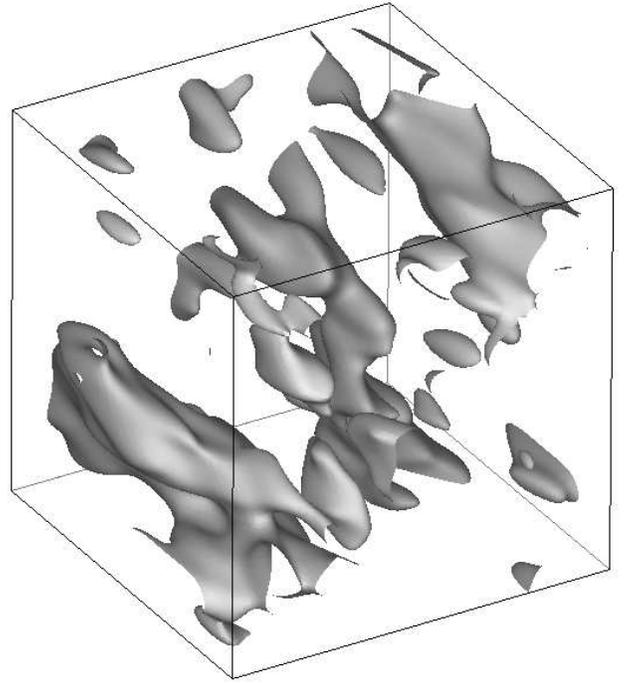
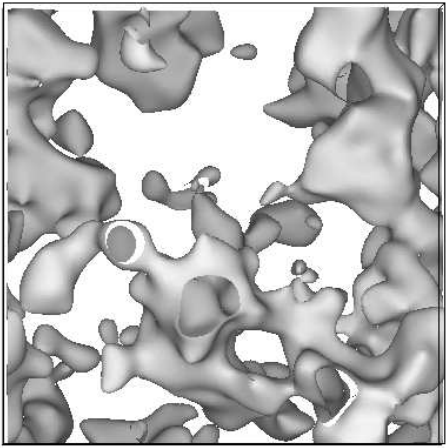
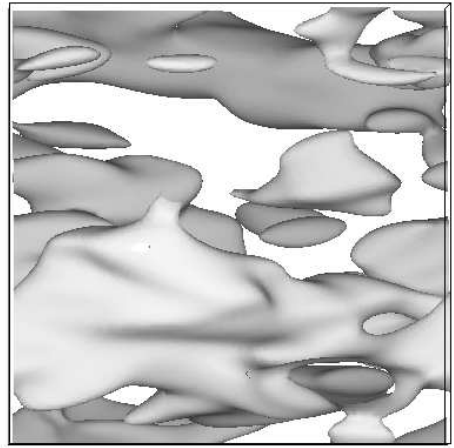
\end{document}